\documentclass[12pt,a4paper]{article}                                                    % previous: 11pt
\usepackage{amsmath, amsthm, amssymb, pdfsync, psfrag, verbatim,color, doi,array}     % add eulervm, doi 
\usepackage[T1]{fontenc}
\usepackage{authblk}
\usepackage {mathrsfs, graphicx, newcent,wrapfig,multirow}
\usepackage{breakcites}
\usepackage{longtable}
\usepackage{caption}
\usepackage{subcaption}
\usepackage[round, square,sort,comma,numbers]{natbib}                                    % add [square, sort, comma, numbers]                       
\newcommand{\bbE}{\mathbb E}

\newcommand{\bbP}{\mathbb P}

\newcommand{\bbR}{\mathbb R}

\newcommand{\crl}[1]{\ensuremath{ \left\{ #1 \right\} }}

\newcommand{\brak}[1]{\ensuremath{\left( #1 \right)}}

\newtheorem{theorem}{Theorem}[section]
\newtheorem{definition}[theorem]{Definition}

\newtheorem{remark}[theorem]{Remark}
\newtheorem{example}[theorem]{Example}
\newtheorem{examples}[theorem]{Examples}
\newtheorem{foo}[theorem]{Remarks}
\newenvironment{Example}{\begin{example}\rm}{\end{example}}

\newenvironment{Remark}{\begin{remark}\rm}{\end{remark}}

\title{Measures of Reliability Risk for\\the Australian Energy Sector}
\date{\today}

\author[1]{John Boland}
\author[2]{Matthias Fresacher}
\author[3]{David Hill}
\author[4]{Shijia Jin}
\author[4]{Ya Li}
\author[5]{Graham Mills}
\author[4]{Kihun Nam}
\author[5]{Craig Oakeshotte}
\author[6]{Andriy Olenko}
\affil[1]{UniSA STEM, The University of South Australia, Australia}
\affil[2]{School of Computer, Data and Mathematical Sciences
Western Sydney University, Australia}
\affil[3]{Department of Electrical and Computer Systems Engineering\\ Monash University, Australia}
\affil[4]{School of Mathematics, Monash University, Australia}
\affil[5]{Australian Energy Market Commission}
\affil[6]{Department of Mathematical and Physical Sciences, La Trobe University, Australia}

\begin{document}
	\maketitle
\begin{abstract}
    The article identifies the desired properties of the reliability risk metric for the Australian Energy Sector. It proposes a comprehensive set of metrics designed to measure various aspects of the reliability risk when a large portion of power generation is composed of variable renewable energy (VRE). The suggested methods aim to balance the tradeoff between economic benefit and risk management, effectively address the tail risk, and interpret the severity of the outage in a direct way. Properties of the considered reliability metrics are investigated by using simulation studies that use time series of expected unserved energy values. Sample distributions, dependency structures and other statistical properties of the metrics are studied. The results suggest new tools and approaches that can be used to increase the efficiency of the Australian Energy Sector.
\end{abstract}

\section{Introduction}
%This section of the paper presents a concise overview of the background and rationale for the investigated problem.

The power sector is currently undergoing a significant transformation driven by the need to address climate change, reduce greenhouse gas emissions, and transition towards a sustainable energy future. This transformation will bring numerous benefits by integrating a higher share of renewable energy sources, such as wind and solar, into the electricity grid. However, it also presents several new challenges for maintaining the reliability and resilience of the power system.

Renewable energy is a critical component of our energy mix consumption. Given the anticipated increase in our reliance on these power sources over time, it is essential to establish a new standard of reliability that accounts for the variable renewable energy (VRE), which depends on the weather conditions \cite{mancarella2022considerations}.

Currently, the metric employed to evaluate the reliability of energy generation is the expected unserved energy (expected USE). However, the expected USE is not designed to account for VRE power generation because they are heavily correlated to each other. This highlights the necessity of a new reliability standard that can assess the ‘tail risk’ of power generation. Another previously proposed approach was a linear combination with conditional value-at-risk (also known as expected shortfall).

To extend the used approaches and to address some of the above challenges, this paper proposes a comprehensive framework for new reliability standards that have been specifically designed to address the unique characteristics of the evolving power sector. These standards have been developed through an extensive discussion conducted by the Australian Energy Market Commission and involving academics.

%\subsection{The Changing Power System Landscape}
The power system is transitioning from a primarily capacity-limited thermal power system to a more energy-limited high VRE power system. This shift necessitates a reevaluation of the existing reliability standards to ensure they remain adequate for the future National Electricity Market (NEM). The existing standards were designed for a previous era and may not properly address the challenges posed by the integration of renewable energy sources and changing demand patterns (for example, discussions in \cite{Barradale, Bigerna, Eryilmaz, Can, Shafiullah, Haugen}).

%\subsection{The Need for New Reliability Standards:}
The proposed new reliability standards aim to enhance the resilience of the power sector by providing a robust framework for assessing and ensuring resource adequacy in the face of increasing variability and uncertainty. These standards will enable effective system planning, recommendations for participants and regulations. They will also facilitate the integration of renewable energy sources, and mitigate the risks associated with potential disruptions.

The primary objective of this paper is to present a comprehensive proposal for new reliability standards to be aligned with the evolving power system. The proposed standards will address the gaps in the current approaches to assess reliability. We identify the desired properties of the reliability risk metric and propose a list of different metrics to address the reliability risk when a large portion of power generation is composed of VRE. A new reliability risk metric should balance the tradeoff between economic benefit and risk management, effectively address the tail risk, and be interpreted to the severity of the outage in a direct way.

First we introduce mathematical notations, which will be used to formalise and describe the main problem and the proposed approach in a rigorous quantitative manner. Let us denote by  $X_t$ the USE, which represents  the electricity power demand minus the gross energy generation at time $t.$ The reliability risk event occurs when $X_t>0$.  The general framework for choosing the reliability risk metric aims to minimise the combined expenses of investment, operation, and consumer costs. The consumer cost will be determined by a function of reliability risk metric, denoted as $\rho(X),$ which depends on the path of $X_t$.

Such metric $\rho$ can be obtained through a combination of selected risk measures, time-dependency functions, and state transformations. Essentially, an infinite number of choices are available, see, for example, the Chen and Cheng \cite{chen2022measuring} discussion on measuring tail risks. In this paper, we focus on presenting a selection of significant choices for each function to construct meaningful reliability standards.

Section~\ref{sec2} develops an approach to define associated risk metrics based on modelling of the USE via stochastic processes. The reliability standards in this formulation are specific functionals of random processes. Several subsections discuss the selection of  components within this general stochastic framework for practical applications, ensuring the fulfilment of desirable properties. Then, Section~\ref{sec3} presents various numerical studies. It illustrates the general concepts and investigate some of properties of the suggested metrics via simulation studies. In the concluding Section~\ref{sec4} we provide a summary of the study's findings, address its limitations, and highlight new research questions that have arisen as a result of this investigation.

\section{Modelling Reliability Standards}\label{sec2}

\subsection{Notations and Definitions}
In this subsections, we provide some general mathematical notations that will be used in the following subsections to introduce the problem, model, and suggested approaches in a rigorous quantitative way.

Let $\bbR$ denote the set of real numbers. To avoid unnecessary complicity, we will only consider a finite time horizon $[0,T]$ where $T$ is large enough to include the time horizon of interest. Unless otherwise noted, the time unit will be a year.  The symbols $c$ and $\gamma$ are used to represent generic finite constants, which are not necessarily the same in each appearance. The notation ${\bf 1}_{x \geq c}$ denotes the indicator function of the event $x \geq c,$ which takes the value $1$, when the event happened, and 0 otherwise.

% All the spaces defined in this paper will be equipped with appropriate norms and $\sigma$-algebras to ensure the well-posedness of the mathematical objects introduced and subsequently used.

Let $(\Omega, \mathcal{F}, \mathbb{P})$ denote a probability space, where the measure $\mathbb{P}$ represents the quantification of our belief in the underlying model. This probability is employed for computations in all theoretical constructions of the model and the corresponding risk metrics. In the context of real data applications, the empirical probability derived from the data is used and will be referred to as $\hat{\mathbb{P}}$. It is assumed that the probabilities $\mathbb{P}$ and $\hat{\mathbb{P}}$ exhibit closeness to each other according to some metric (e.g., the Wasserstein distance defined in \cite{kantorovich1960mathematical} and \cite{vaserstein1969markov}) defined on probability measures.

We will denote $L^0(A;B)$ to be the set of measurable
%\footnote{Measurability in this paper is a technical condition to define mathematical concepts accurately.}
functions from $A$ to $B$. For example, $L^0(\Omega;\bbR)$ is the set of random variables. Also, we will denote $C([0,T];\bbR)$ be the space of continuous functions from $[0,T]$ to $\bbR$, equipped
% with the Borel sigma algebra
with the sup norm. For example, $L^0(\Omega; C([0,T];\bbR))$ is the space of measurable functions with values in continuous paths from $[0,T]$ to $\bbR$.

The risk event of concern is the case where the supply of electricity does not meet the demand. In order to measures the severity of the risk, we use risk metrics. The main data that will be used for the risk metrics is called USE, denoted by $(X_t)_{t\in[0,T]}$. They are given as the paths of the difference between unserved power and the total extra capacity of generators. Therefore, the desirable scenario is  $X_t\leq 0$, which means the supply of electricity meets the demand. On the other hand, $X_t<0$ is a risk event. This quantity is closely related to the conventional Unserved Energy (cUSE). Note that the cUSE measures the unserved power when the generators are running at full capacity and 0 when the generators are not running at their maximum. Therefore, cUSE is always nonnegative. When the generators are running at their maximum, USE would be equal to cUSE; while they are not, USE would be the negative of the generators' remaining capacity. Unless otherwise noted, we will use the notation $X$ for the USE and $X_t:=X_t(\omega)$ for the corresponding random variable at time $t$. We will assume that the USE $(X_t)_{t\in[0,T]}$ is a real-valued continuous stochastic process and its current and historical values are always accessible.

\subsection{Concept of Reliability Standard Metric}
%This subsection discusses some desirable properties of a reliability standard.
The reliability standard should be a functional of USE.  Usually it is applied to evaluate future risk of electricity shortfall or perform retrospective evaluation based on historical data. In the first case it will be applied to simulated scenarios, in the second case, to a particular realisation of the USE.

For practical applications and interpretability, it is desirable that a reliability standard has the monotonicity properties. The following is the rigorous definition.

\begin{definition}
	The reliability standard $\rho$ is a functional
	\[
	\rho: L^0(\Omega; C([0,T];\bbR))\to\bbR
	\] that satisfies the following properties:
	\begin{itemize}
		\item[] (monotonicity) If $X_t\leq Y_t$ for all $t\in[0,T]$, then
		$\rho(X)\leq \rho(Y)$;
		\item[] (normalised) $\rho(0)=0$.
	\end{itemize}
\end{definition}
\begin{Remark}
    For simplicity, let us assume that the regulatory standard is given as a single reference value (in the real situation, it can also be a vector of such values) denoted by $c\in\bbR.$

	The reliability standard measures the risk of electricity shortfall to the demand, that is the USE. For example, for a given USE, $X$, and the regulatory standard, $c,$ if $\rho_1(X)\leq c\leq\rho_2(X)$, then one can interpret the $\rho_1$ as a standard satisfying the regulatory standard, but $\rho_2$ does not. In this sense, if $\rho_2$ is the higher reliability standard and provides a higher standard for the regulatory viewpoint.
\end{Remark}
\begin{Remark}
	The definition of the reliability standard is similar to the risk metric used in finance: see \cite{artzner1999coherent}
\end{Remark}

It is important to note that not all reliability standards are  suitable for specific practical applications. The fitness of candidate standards should be assessed based on their capacity to reflect  an efficient level of reliability and identify the overall efficient portfolio of power system resource investments, taking into account the risks to reliability. For example, the reliability standard is desired to  effectively restrict the duration, magnitude, and frequency of specific USE risky events to bounds that are considered acceptable (within an appropriate level of confidence).  Additionally, it should enable explicit valuation of investments to address high-impact low probability events (tail risk).

The following subsections propose a set of basic reliability standards and outline a methodology to construct more complex reliability standards from the existing ones.

\subsection{Basic Reliability Standards}
Let us denote the USE as $X=(X_t)_{t\in[0,T]}$. In this section, we will construct the reliability standards $\rho$ as a composition of the three basic risk functions: a real-to-real function $\sigma$, a functional $\tau$ which maps a path to a real number, and a functional $R$ that maps a random variable to a real number. More precisely, we define $\rho$ based on the following formula:
\begin{equation}\label{rmain}
\rho(X)=(R\circ\tau)\brak{\brak{\sigma(X_t)}_{t\in[0,T]}}-(R\circ\tau)\brak{\brak{\sigma(0)}_{t\in[0,T]}},\end{equation}
where
\begin{equation}\begin{aligned}
	R&:L^0(\Omega;\bbR)\to\bbR ,\\
	\tau&:L^0([0,T];\bbR)\to \bbR ,\\
	\sigma&:\bbR\to\bbR.
\end{aligned}
\end{equation}
Here, $\tau$ represents time sensitiveness, and $\sigma$ represents the state sensitiveness of the reliability risk, respectively. The functional $R$ is defined on random variables and measures the risk. $R$ can be thought of as a risk measure with opposite monotonicity. That is, a larger random variable represents higher risk, while a classical risk measure takes lower value for a smaller random variable. The discussion revolves around the properties that each of these functions can assess within the USE. The assumptions on these three basic functions and their specific examples  are provided in the following. 

We assume that $\sigma$ is a nondecreasing function of a real argument, and  $\tau$ and $R$ are monotone functions, such that
\begin{itemize}
    \item $X_t\leq Y_t,$ for all $ t\in[0,T],$ implies $\tau(X)\leq \tau(Y),$
    \item	$Z_1(\omega)\leq Z_2(\omega),$  for all $\omega\in\Omega,$ implies $R(Z_1)\leq R(Z_2),$
\end{itemize}
for any $X,Y\in C([0,T];\bbR)$ and random variables $Z_1$ and $Z_2.$ These assumptions guarantees the monotonicity of $\rho$.

The reliability risk in (\ref{rmain}) can be constructed via a specific combination of $R,\tau,$ and~$\sigma$. For example, if one uses $\tau(x):=\int_0^TX_tdt$ and $\sigma(a)=\max(a,0),$ then the corresponding reliability standard would be time-homogeneous (no specific time is more sensitive to the reliability risk). Additionally, it is ignorant of the scenario when the demand is less than the capacity ($X_t<0$). In particular, if one takes $R$ as an expectation, this becomes the conventional Unserved Energy. One can also take $R$ as the mean-CVaR, then it becomes the reliability risk recently suggested by Mancarella \cite{mancarella2022considerations}.

In the following subsections, we will introduce and discuss several examples of the functions $R,\tau,$ and $\sigma$. Depending on the context, one can choose an appropriate combination to represent a reliability risk with desirable properties.
\subsubsection{Choices for $R$}
The suggested metric $R$ acts on random variables. In the following, $Z$ denotes a random variable, $f_Z$ and $F_Z$ are its probability density and cumulative distribution functions respectively. We suggest to use the following functions $R,$ which are widely employed in the literature as risk measures for random factors:
\begin{itemize}
	\item[(R1)] $R(Z)=\bbE Z,$ i.e. $R$ is just the expected value. This operator is linear and easy to understand. However, it is well-known, that it cannot incorporate the tail risk effectively.
	\item[(R2)] $R(Z)=\bbP(Z\geq c) = 1- F_Z(c)$ for a given $c\geq 0$. This $R$ represents the right-tail probability of $Z$. It is a descriptive metric and also easy to understand and implement. However, it lacks the property of positive homogeneity, that is, $R(kZ)\neq kR(Z)$. Consequently, the selection of the scale could influence risk control, potentially relying on subjective opinions.
	\item[(R3)] $R(Z)=\frac{1}{\alpha}\int_0^\alpha F_Z^{-1}(1-x)dx$ for a given $\alpha\in(0,1)$. This quantity is essentially the negative Conditional Value-at-Risk for the random variable $-Z$. It measures the right-tail risk of the random variable $Z$. Such $R$ was well-studied and widely used in the finance industry due to its coherency (that is, it is scale-blind, and the diversification reduces risk) while incorporating the tail risk.
	\item[(R4)] $R(Z)={\arg\max}_{x\in\bbR} f_Z(x).$ This $R$ measures the most likely value for $Z$. Similar to the case (R1), this metric does not incorporate the tail risk.
\end{itemize}
\begin{Remark}
	It should be noted that while the measures in (R1) and (R4) do not directly incorporate the tail risk information, we can deal with the tail risk by appropriately selecting the two other functions $\tau$ and $\sigma$ in (\ref{rmain}).
\end{Remark}

\subsubsection{Choices for $\tau$}
The suggested metric $\tau$ acts on trajectories of random processes $X.$  We suggest to use the following functions $\tau,$ which are widely employed in the literature on functionals of stochastic processes:
\begin{itemize}
	\item[(T1)] $\tau(X)=\int_0^Tw(t)X_t\,dt$ for some non-negative function $w:[0,T] \to[0,+\infty).$   This functional is the weighted average of the observed values, where $w(t)$ represents the weights that can be used to reflect varying contributions of observations across different time periods. For instance, to account for periodic changes, one can consider $w(t) = 2 - \sin(2\pi t)$. This form of $\tau$ can be employed to incorporate a time-varying sensitivity to reliability risk.
	\item[(T2)] $\tau(X)=\sup_{t\in[0,T]}X_t$. This $\tau$ gives the maximum risk during the given time horizon~$[0,T].$
    	\item[(T3)] $\tau(X)=\int_0^T{\bf 1}_{X_t\geq 0}dt,$ where ${\bf 1_A}$ is an indicator function of an event $A.$ This functional $\tau(X)$ provides the total duration time of outages  during the time period $[0,T].$
	\item[(T4)] $\tau(X)=\sup\crl{\gamma\in\bbR:\int_0^T{\bf 1}_{X_t\geq \gamma}dt\geq cT},$ where $\gamma$ and $c\in [0,1]$ are given  constants. For a given value of $c$, the function $\tau(X)$ provides the highest threshold $\gamma$ such that the duration of time that the process $X$ surpasses it is at least $cT.$
\end{itemize}
For given $X\in L^0([0,T];\bbR)$, let us define the following times
\begin{equation}\begin{aligned}
	\upsilon^s_1(X)&=\inf\crl{t>0:X_t>0},\\
	\upsilon^e_1(X)&=\inf\crl{t>\upsilon^s_1(x):X_t< 0},\\
	\upsilon^s_n(X)&=\inf\crl{t>\upsilon^e_{n-1}(x):X_t>0},\\
	\upsilon^e_n(X)&=\inf\crl{t>\upsilon^s_n(x):X_t< 0}.
\end{aligned}
\end{equation}
These values represent the time instances of consecutive crossing of level 0 from either above or below. Therefore, $[\upsilon_i^s(X), \upsilon_i^e(X)]$ is the time interval for $X$ being greater than $0$. In particular, $[\upsilon_i^s(X), \upsilon_i^e(X)]$ can be interpreted as the $i$th outage event. Using these notations, one can define the following metric $\tau.$
\begin{itemize}
	\item[(T5)] $\tau(X)=\sup\crl{n\geq 0:\upsilon^s_n(X)<T}.$ This $\tau$ gives the number of time intervals where $X$ exceeds $0$. Thus, $\tau(X)$ can be used to compute the number of outages during $[0,T].$
\end{itemize}
As it was already mentioned, these functionals can be used to evaluate future risk by applying them to simulated scenarios or to perform retrospective evaluation via applications to historical data.  It should be noted that in real applications the values of the random process $X$ are sampled only at discrete time moments. Consequently, when applying the aforementioned functionals to real data, one should use the discretised versions of the functionals over the observation time grid and replace integrals with sums.

\subsubsection{Choices for $\sigma$}
The following functions $\sigma$ provide scaling and threshold transformations for the USE, which can be suitable in various practical situations. For instance, these transformations can be used to incorporate electricity storage capacity, outages that exceed a specific level and other similar scenarios.
\begin{itemize}
	\item[(S1)] $\sigma(a)=\max\crl{a,0}$. As previously explained, this specific $\sigma$ assumes that there is no risk associated with $X_t<0,$ i.e. that  negative values of $X$ do not diminish the risk. This assumption is applicable in situations where electricity storage options, such as batteries, are unavailable.
	\item[(S2)] $\sigma(a)=ca,$ where $c$ is a positive constant. Here, $c$ can be interpreted as the sensitivity with respect to the demand minus capacity. Since $X_t<0$ lessens the reliability risk accumulated by $X_t>0$ events when the time-integral metric $\tau$ is used,  this particular $\sigma$ is suitable in scenarios where substantial electricity storage, such as batteries, is present.
	\item[(S3)] $\sigma(a)=c\max\crl{a,-\gamma},$ where $c$ and $\gamma$ are positive constants. This is a version that combines the above two types of $\sigma.$ Combined with $\tau$ being time-integral, $c$ represents the sensitivity with respect to the demand minus capacity, and $\gamma$ represents the electricity storing capacity.
	\item[(S4)] $\sigma(a)={\bf 1}_{a>c}$ for a given constant $c.$
    This function indicates whether the USE surpasses the threshold $c$. Specifically, when $c=0$ is set, it encodes information regarding the occurrence of an outage (demand exceeds capacity).
\end{itemize}
\subsubsection{Constructing the Reliability Standard}
 Finally, to construct the reliability standards one can select and substitute the considered  functions $R, \tau$ and $\sigma$ in the equation (\ref{rmain}).  Table~\ref{examples} provides some examples of important combinations of appropriate $R, \tau$ and $\sigma$ along with their respective advantages and disadvantages.  The first two metrics are the currently used one and recently proposed by Mancarella \cite{mancarella2022considerations}.
  Note, this table only contains a subset of 8 out of 80 possible superpositions of the functions $R,\tau,\sigma$ listed above.
\begin{center}
	\begin{longtable}{>{\raggedright}p{3cm}|p{2.5cm}|p{7cm}}
	% \begin{longtable}{|p{4cm}|p{3.5cm}|p{7cm}|}
            \caption{Metrics $\rho$ as Superpositions of  $R, \tau$ and $\sigma$}\label{examples}\\
		\hline
		\textbf{Metric} & \textbf{$R,\tau,\sigma$} & \textbf{Advantages and disadvantages} \\
		\hline 
		Mean of Conventional Unserved Energy & (R1)(T1)(S1) & Used in the Current Relibility standard. Simple to understand. The risk associated to extreme events may be underrepresented.\\
            \hline
		Conditional Value at Risk (CVaR) & (R3)(T1)(S1) & (See in \cite{mancarella2022considerations}) In wide use for fat tail distributions. No information on the extent of the tail, its shape or extremes. Not easy to understand. \\
		\hline
		Most Probable Maximum Risk (MPMR) & (R4)(T2)(S2) & An estimate of maximum unserved energy that would be expected to occur in a single event. As it is the mode of the distribution of maxima from subsets, it may not incorporate all information from the tail of the distribution. Also, it may not provide insight into the customer experience with respect to frequency or duration. \\
		\hline
		Expected Maximum Risk (EMR) & (R1)(T2)(S2) & More realistic than MPMR. Uses an expected value of the maxima rather than an estimate of the possible maximum. It could take very high values if the tail becomes long. \\
		\hline
		Tail probability of the maximum of the USE & (R2)(T2)(S2) & Deals with a measure of maximum USE. Selected reference value of USE could be arbitrary, and then it  provides a probability that this or larger USE values occur. \\
		\hline
		Quantile of conventional Unserved\newline Energy & (R2)(T1)(S1) & Simple and descriptive. The used scaling can be subjective. \\
		\hline
		Mean of outage total duration & (R1)(T1)(S4) & Intuitively important. Requires determining of the sensitivity of consumers to different outage durations. \\
		\hline
		Mean frequency of outages  & (R1)(T5)(S2) & Intuitively important. Inexplicitly connected to outages duration. It is unclear whether a prolonged outage is worse than multiple shorter outages with an equal total duration. Additionally, the impact of the time pattern in which these outages occur remains unclear.  \\
		\hline 
	\end{longtable}
\end{center}
\begin{Remark} Note that some of the metrics possess a characteristic (convexity) that result in lower risk when the energy supply becomes diversified. This property can be desirable in practical situations.
\end{Remark}

\subsection{Combination of Reliability Standards}
More complex reliability standards can be employed to achieve finer control over various aspects of energy market reliability. Specifically, one can simultaneously utilize multiple standards, denoted as 
$\rho_1(X)$, $\rho_2(X)$, $\ldots$, $\rho_n(X),$ where $\rho_i$ represents the reliability standards mentioned in the previous section. In such cases, the establishment of specific constraints for each reliability standard results in the verification of a set of multivariate inequalities.

 For example, consider the function \begin{align}\rho(X)=(\rho_1(X), \rho_2^\alpha(X)),\quad \alpha \in(0,1),\end{align} where $\rho_1$ is the expected USE, $\rho_2^\alpha$ is the $\alpha$-conditional value-at-risk, and the admissible set is the half-plane of the $xy$-plane. In this case, the method corresponds to the mean-CVaR reliability standard. Specifically, if one defines the half-plane as the left side of the $xy$-plane, it represents the current reliability standard based on expected USE.

Alternatively, an approach involving the joint combination of various standards can be adopted. In this case, the relaxation of certain standards imposes more stringent restrictions on the remaining ones. For instance, one may opt for a simple average or a weighted average of the $\rho_i$ values, assigning equal or varying degrees of credibility to the included standards based on their respective weight values. It can be seen that this scenario employs a linear multivariate function.

In more general settings, let us consider a function $G:\bbR^n\to\bbR\cup\crl{\infty}$, which is non-decreasing in each variable  Then, it is easy to see that the function
\begin{align}\rho(X)=G(\rho_1(X),\rho_2(X),...,\rho_n(X))-G(0,0,...,0)\end{align}
is again a reliability standard.

Note that, in general, the set of $\rho_i$s can contain repeated values, which can be used to define several restrictions with respect to the same standard $\rho_i.$

\begin{Example}
    The mentioned above weighted linear combination of  reliability standards can be expressed as
\begin{align}
    G(r_1,r_2,...,r_n)=\sum_{i=1}^{n}w_ir_i,
\end{align}
   where  $(w_i)_{i=1,2,...,n}$ are constant  weights.
 \end{Example}

\begin{Example}
	For a given $k\leq n$, admissible criteria $c_{k+1},c_{k+2},...,c_n$, weights $(w_i)_{i=1,2,...,k}$ and a constant $\gamma\geq 1$, let
	\begin{equation}\label{wsum}
	G(r_1,r_2,...,r_n)=\brak{\sum_{i=1}^{k}w_ir_i^\gamma}^{1/\gamma}+\brak{\prod_{i=k+1}^n{\bf 1}_{r_i\leq c_i}}^{-1}.
	\end{equation}
	The first term on the right-hand side of (\ref{wsum}) is the weighted $l^\gamma$-norm of a vector. When $\gamma=1$, it is a linear combination of the reliability risks we consider. As $\gamma$ increases, the reliability risks with more extreme values contributes higher to the reliability standards.

	The second term on the right-hand side of (\ref{wsum}) defines the admissible set of USE. For example, if one of $\rho_i(X)$ for $i=k+1,k+2,...n$ exceeds the corresponding admissible criterion $c_i$, this term would give infinite value resulting $\rho(X)$ to be infinite. Therefore, it will violate the regulatory criterion. In that sense, the second term ensures $\rho_i(X)\leq c_i$ for all $i=k+1,...,n$.
\end{Example}

The tradeoff between different reliability risks can be modelled by other complex functions similar to the examples provided above. This modelling  should rely on a careful analysis of data from power suppliers and customer surveys.
\subsection{Application of a Reliability Standard to\\ the Energy Market}
This section concludes the methodological part of the paper by outlining and clarifying the role of the proposed reliability standard on the energy market.

The application of the reliability standard to the energy market can be understood through the following optimisation problem.
Let us consider a portfolio of generators and batteries and denote it by the symbol $\pi$. The investment cost associated with this portfolio will be denoted by $Inv(\pi)$, while the operational cost will be denoted by $Op(\pi)$.

The stochastic process $(X_t^\pi)_{t\in[0,T]}$ represents the maximum power for the portfolio minus electricity demand at time $t$. The function $P_1$ describes the penalty in case the reliability standard is violated, and the function $C$ gives the consumer cost for a given $X^\pi$. Additionally, the function $P_2$ gives the penalty/cost for choosing a specific reliability standard.

Now, for example, consider the case when the penalty $P_1$ is applied if the value of the reliability standard $\rho(X^\pi)$ exceeds a specific threshold $\alpha$.
Under these notations and assumptions, the cost function associated with a specific portfolio, the USE process $X^\pi$, and the reliability standard $\rho$ is given by $Inv(\pi) + Op(\pi) + C(X^\pi) + P_1({\bf 1}_{\rho(X^\pi) > \alpha}) + P_2(\rho)$.

Therefore, for a given reliability standard $\rho$ the optimal investment can be obtained as the solution to the following optimisation problem
\begin{align}\min_{\alpha, \pi} (Inv(\pi) + Op(\pi) + C(X^\pi) + P_1({\bf 1}_{\rho(X^\pi) > \alpha}) + P_2(\rho)).\end{align}
For simulated $X^\pi$ and various choices of $\rho$, the resulting costs and optimal values can serve as indicative benchmarks for both market participants and the regulator.

\section{Empirical Studies}\label{sec3}
This section illustrates some of the discussed concepts. The explanations and examples have been kept simple to ensure accessibility and transparency for energy market practitioners and regulating authorities.

\subsection{Data}
The 2022 Electricity Statement of Opportunities (ESOO), prepared for the Australian Energy Market Operator gives forecasts of electricity supply reliability in the NEM.  The metric that is generated is USE.  The forecasts were done for the ten-year period 2022-23 to 2031-32.  From the ESOO, the methodology used is described as

"Applying a statistical simulation approach which assesses the ability of existing and committed generation to meet forecast demand in all hours. The model calculates expected USE over a number of forecast conditions impacting demand and renewable generation (based on 11 historical reference years of weather) and random generator outages, weighted by likelihood of occurrence, to determine the probability of any supply shortfalls. These shortfalls have been expressed in terms of the forecast expected USE."

For the purposes of illustration of some of the risk metrics described in the paper, we use the forecasts of USE for NSW for the period 1/7/2025 to 30/6/2028.

\begin{figure}[!ht]
    \centering
    \begin{subfigure}[b]{0.49\textwidth}
        \centering
        \includegraphics[width=\textwidth]{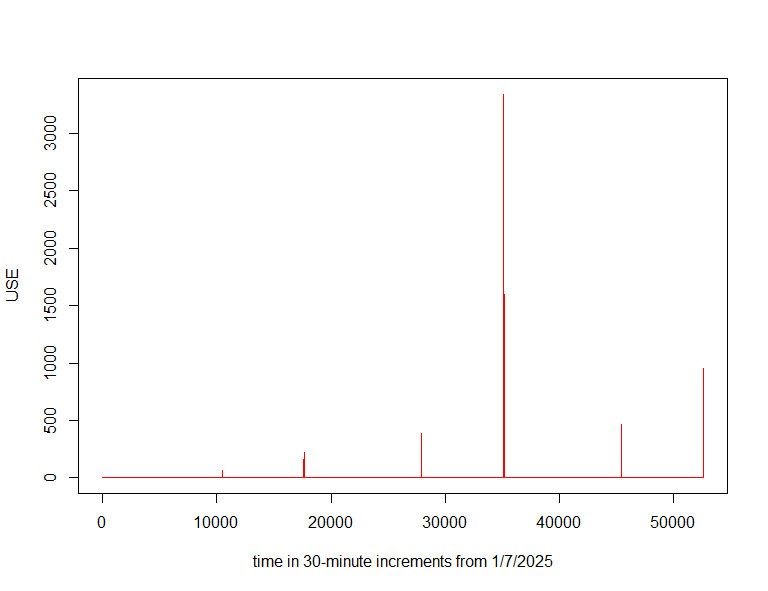}
        \caption{Full trajectory}
        \label{f0a}
    \end{subfigure}
    \hfill
    \begin{subfigure}[b]{0.49\textwidth}
        \centering
        \includegraphics[width=\textwidth]{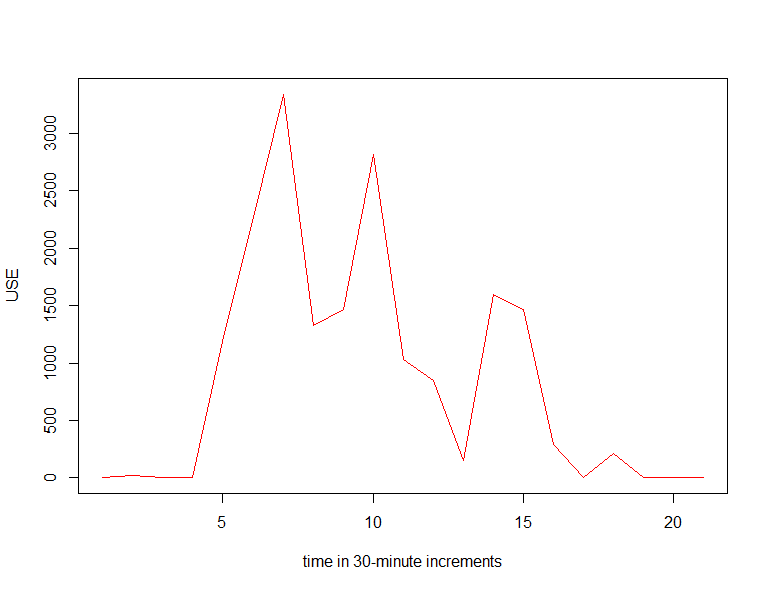}
        \caption{Longest outage}
        \label{f0b}
    \end{subfigure}
    \caption{Simulated USE realisation}
\end{figure}

In Figure~\ref{f0a} you can see a plot of one of the simulated trajectories of USE values with a time lag of 30min. The second subplot,  Figure~\ref{f0b},  shows a part of the trajectory zoomed in around the longest outage.
\begin{figure}[!htb]
    \centering
    \includegraphics[width=0.9\linewidth]{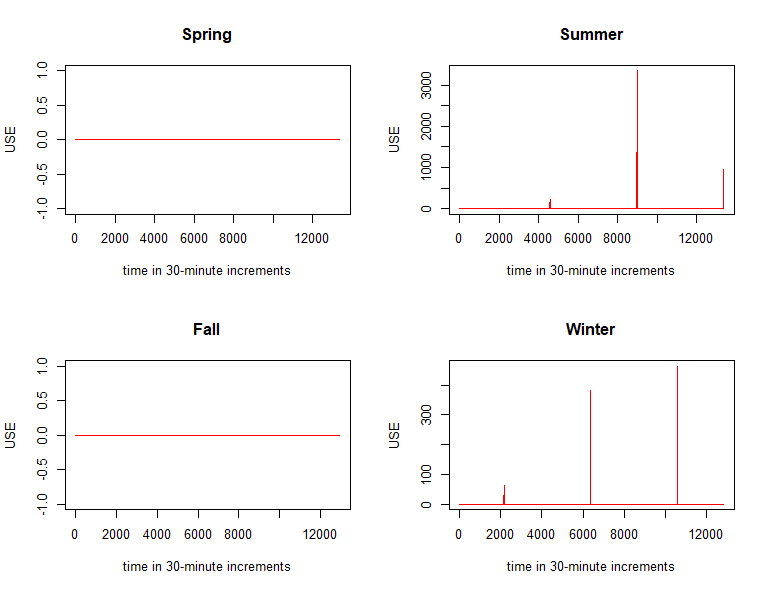}
    \caption{USE trajectories grouped by seasons}
    \label{f1}
\end{figure}

Figure~\ref{f1} illustrates the USE realisation values categorised by their corresponding seasons. The plots demonstrate that the ESOO's model, employed for the simulations, yielded no outage cases during the spring and autumn periods. However, there were some outages in the winter and summer periods, with significantly higher values occurring during the summers.

\subsection{Estimates of Reliability Metrics}
This subsection presents the estimated values of some of the metrics discussed earlier for the USE data from the previous subsection. These estimates are illustrated by referring to the histogram of NSW USE values.

\begin{figure}[!htb]
    \centering
    \includegraphics[width=0.75\linewidth]{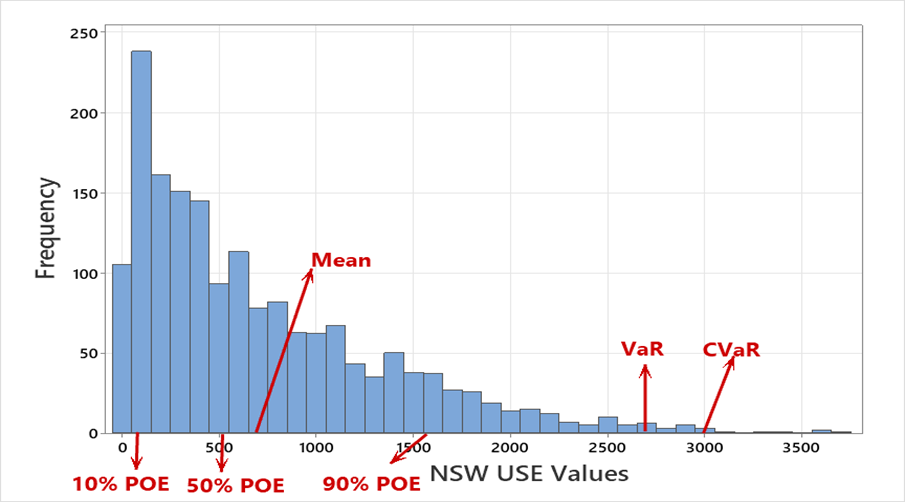}
    \caption{Common Statistical Measures}
    \label{J1}
    % \vspace{-4mm}
\end{figure}

{\bf Pictorial Illustration of Common Statistical Measures:} These terms will be further explained in the document, along with a discussion of risk measures focused specifically on maximal values of USE.  In the diagram shown in Figure~\ref{J1}, POE stands for Probability of Exceedance, VaR is Value at Risk and CVar is Conditional Value at Risk.  To give context, 50\% POE is the Median, the value for which 50\% of the data is below and 50\% above.

\begin{figure}[!htb]
    \centering
    \includegraphics[width=0.75\linewidth]{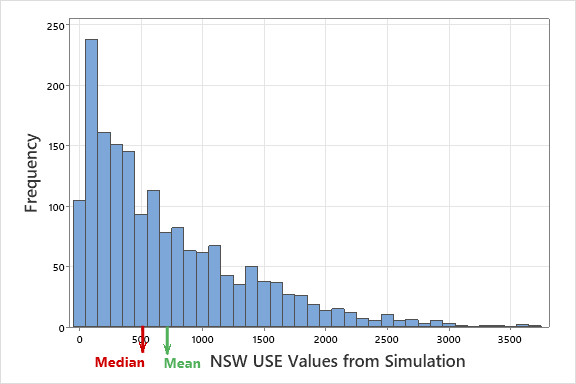}
    \caption{Mean and median}
    \label{J2}
\end{figure}

{\bf Mean of USE:}  This is one of the standard  measures used to evaluate the reliability of NEM.   One of its main issues is that for skewed distributions the mean is not a credible estimate of the central/expected value of the distribution, see Figure~\ref{J2}.  A more fitting approximation is the median, a value where precisely half of the observations are anticipated to fall below, and the other half above it. This is because the mean is overly influenced by extreme values, whereas the median is not. It seems useful in the instance of the NEM data where as the effects of the long tail of the distribution are of significant importance. We will discuss also other metrics that may be more useful metrics to indicate the tail risk. Nevertheless, the median should be used if one is interested in the metric that describes the value of USE that is most typical.  It gives the value of USE that one mostly would expect in a given time period. In the example from the NSW USE simulations, the mean is 706 MWh, while the median is 504 MWh, whose relative positions are shown in Figure~\ref{J2}.

\begin{figure}[!hbt]
    \centering
    \includegraphics[width=0.75\linewidth]{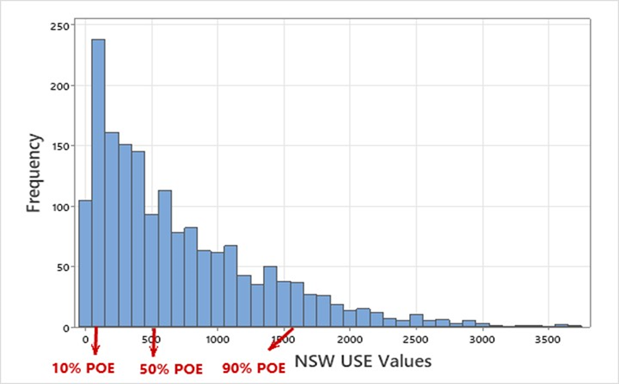}
    \caption{Quantiles of the NSW USE Distribution}
    \label{J3}
\end{figure}

{\bf Quantiles of USE:} Documents from the Australian Energy Market Operator (AEMO) shows that they use as a reference 10\%, 50\% and 90\% Probability of Exceedance (POE) values.  These are synonymous with the suggested 10\%, 50\%, and 90\% quantiles or percentiles. The 50\% quantile corresponds to the considered median.  Similarly, the $\alpha$\% quantile is the value for which one expects that $\alpha$\% of the observations are below that value.
Figure~\ref{J3} gives such values for the example of the NSW USE from the simulations (10\% POE = 72.9 MWh, 50\% POE = 503.8 MWh, and 90\% POE = 1610.0 MWh).  The quantile approach can be applied to the raw values of USE or the distribution of its maxima discussed later.

\begin{figure}[!htb]
    \centering
    \includegraphics[width=0.75\linewidth]{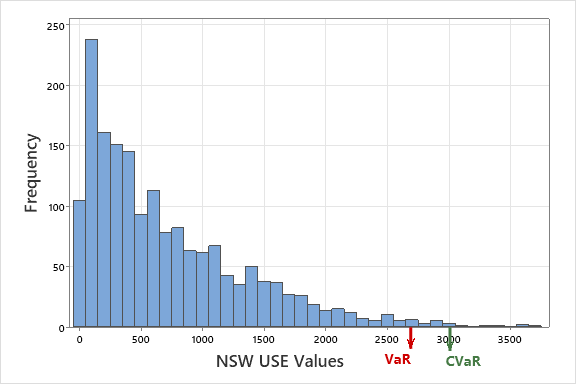}
    \caption{Values of VaR and CVaR for the simulated USE}
    \label{J4}
\end{figure}

{\bf Condition Value at Risk (CVaR):} When discussing Conditional Value at Risk (CVaR), it is better to first define Value at Risk (VaR).  This is a metric commonly used in the risk industry.  The VaR, as depicted in Figure~\ref{J4} for the NSW USE, at a 99\% level is 2,731 MWh.  It means that there is only a 1\% chance that the ‘loss’ will exceed that figure. Unfortunately, VaR is not suitable for optimisation.  Also, as Webby et al. \cite{Webby} and other pulications discussed, it is often the case that the potential loss is very much greater than VaR, even though the probability of this greater loss is very low. This is the case when the distribution has a fat tail.  To cater for these low probability but high-risk events, so-called black swans, an extremely rare event with severe consequences, Rockafellar and Uryasev \cite{Rockafellar} recommend the use of CVaR rather than VaR.  It is simply the mean of the values beyond VaR. Its property as a coherent risk metric includes the ability to easily use it in optimisation. One can easily formulate a problem to minimise the risk as defined by CVaR by adjusting the values of decision variables on which it depends.
As Figure~\ref{J4} shows, in our example, CVaR is 3,079 MWh.  Also, see the relative positions of VaR and CVaR in Figure~\ref{J4}.

{\bf Most Probable Maximum Risk, Expected Maximum Risk and Extreme Value Theory:}
Most Probable Maximum Risk (MPMR) is described in Chen and Cheng \cite{chen2022measuring} as the expected maximum risk (EMR). Thomas et al. \cite{Thomas} provide a description of extreme value theory (EVT).  These will be discussed together as they describe methods for understanding the possible maximum values that a variable can attain.  The rationale is that we select a group of potential extreme values of USE from either historical data or in this case, simulated values from the USE simulations.

There are two methods in the statistical literature for selecting this group of values:
\begin{itemize}
    \item One is the peak over threshold (POT) method.  In this, one has to subjectively select a particular threshold value of USE, say, let us consider 2000 MWh as an example. Then, take all values over 2000 and look at the distributional qualities of this group to make some estimation of a worst-case scenario.
\item The other main method which is often used is the block maximum method.  Here, extremes are created by dividing the analysis period into non-overlapping periods of the same size and then choosing the maximum observation of each new period.
\end{itemize}
Both methods have been used in trying to estimate the maximum amount of rainfall one could expect for a certain location in a given short period of time: see, for example, \cite{bucher2021horse} and reference therein.

\begin{figure}[!htb]
    \centering
    \includegraphics[width=0.75\linewidth]{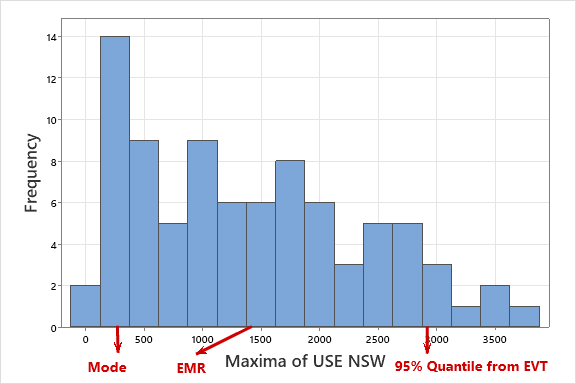}
    \caption{Histogram of the maximum USE values, selected by the block maximum method}
    \label{J5}
    %\vspace{-2mm}  % do not use \vspace
\end{figure}

Now we can clarify the meaning of the metrics for the simulated data for NSW USE. Select the 85 maximum hourly values of USE, which is an example of using the block maxima method.  Figure~\ref{J5} gives the histogram of these maxima.  The MPMR value is given by the peak of the distribution, which is its mode. In this case, it is 375 MWh.  This is probably not much use for the reliability standard.  The EMR is the mean of these maxima, and is given by 1410 MWh.  EVT gives a value of the maximum one could expect at a particular probably level, i.e. in a specific quantile.  For example, the 95\% quantile from the distribution is 3,125 MWh, meaning that only 5\% of the time would one expect that the maximum hourly USE would be over 3,125 MWh.  One has to choose that probability level. For high-reliability scenarios it could be set at 99\% for example.

To compute the probability that the maximum of USE is greater than some particular value one has to use the inverse process from the EVT calculation.  In EVT, one selects a probability and then finds the value in the distribution of the maxima that matches.  In this calculation, one selects a value in the distribution and then finds the probability that values exceed that.

{\bf Outage duration:} Duration means the length of time of any specific outage.  This may be of particular importance as there could exist a threshold of duration above which the impact on consumers changes from a manageable situation to one that involves significant financial loss.
\subsection{Estimates of Metrics Distributions}
The standard visual method for inferring possible values of outages and their likelihood is by using histograms. They are one of the main tools for gaining initial insights into the potential distribution's shape, skewness, tails, extreme values, and other characteristics. They are also employed in assessing reliability metrics and formulating hypotheses about expected outage patterns.

In the previous subsection, we used the histogram of original simulated USE values for illustrative purposes. These simulated USE values can also serve as the basis for deriving other statistics, such as empirical values of the metric $\tau$ or similar metrics related to realisations of USE. While it is usually difficult to derive distributions of other non-linear statistics analytically, it can be easily done by using simulations. This subsection provides several examples of such estimated distributions.

\begin{figure}[!htb]
    \centering
    \includegraphics[width=0.75\linewidth]{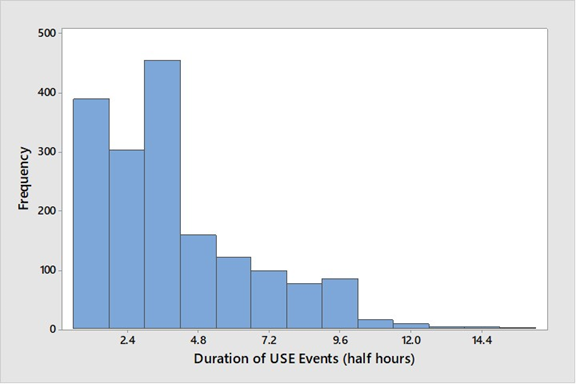}
    \caption{Histogram of the duration of outages}
    \label{J6}
\end{figure}

Referring back to the last part of the preceding subsection, Figure~\ref{J6} presents an example of the estimated distribution for outage duration (in half-hours) in the NSW simulations.
 The way to use this type of display is to estimate the probability of different durations.   For example, there is only a 2\% chance of an outage lasting longer than 10 hours, an 18\% chance of longer than 6 hours, but a 40\% chance that it would be less than 2 hours.

\begin{figure}[!htp]
         \centering
     \begin{subfigure}[b]{0.48\textwidth}
         \centering
         \includegraphics[width=\textwidth]{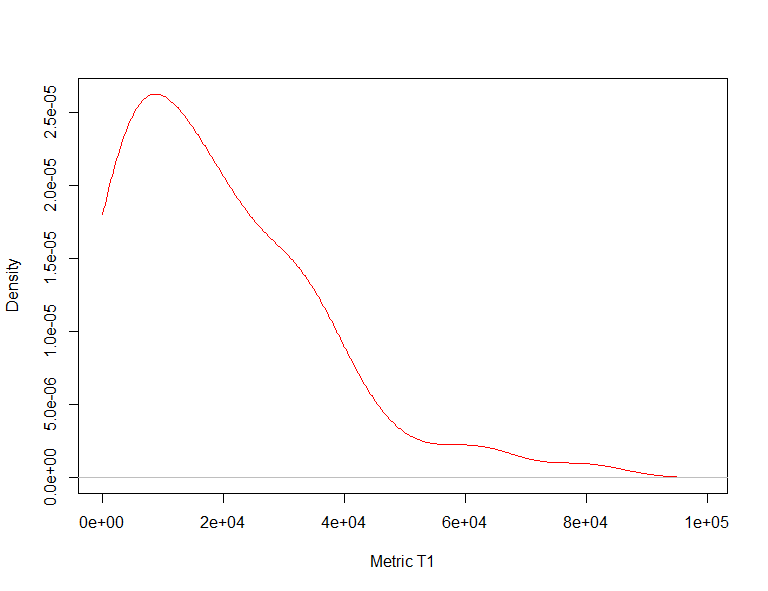}
     \end{subfigure}
          \begin{subfigure}[b]{0.48\textwidth}
         \centering
         \includegraphics[width=\textwidth]{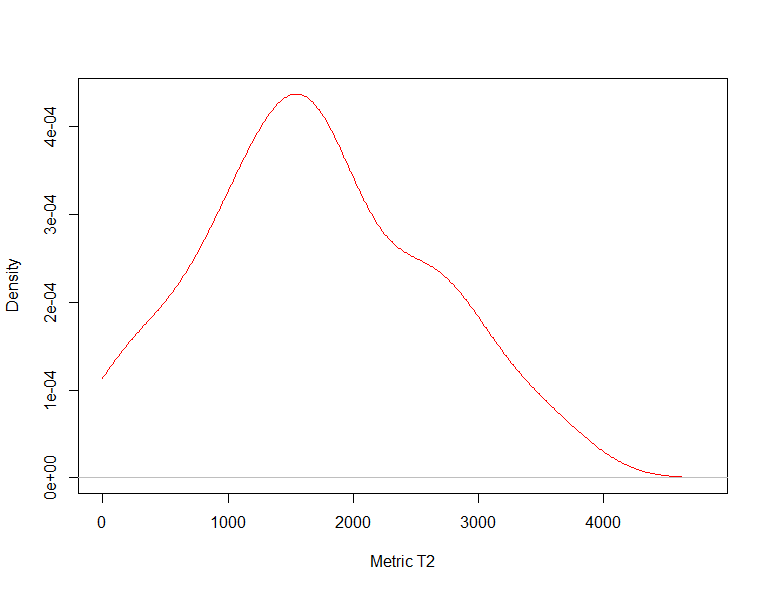}
         \end{subfigure}
              \begin{subfigure}[b]{0.48\textwidth}
         \centering
         \includegraphics[width=\textwidth]{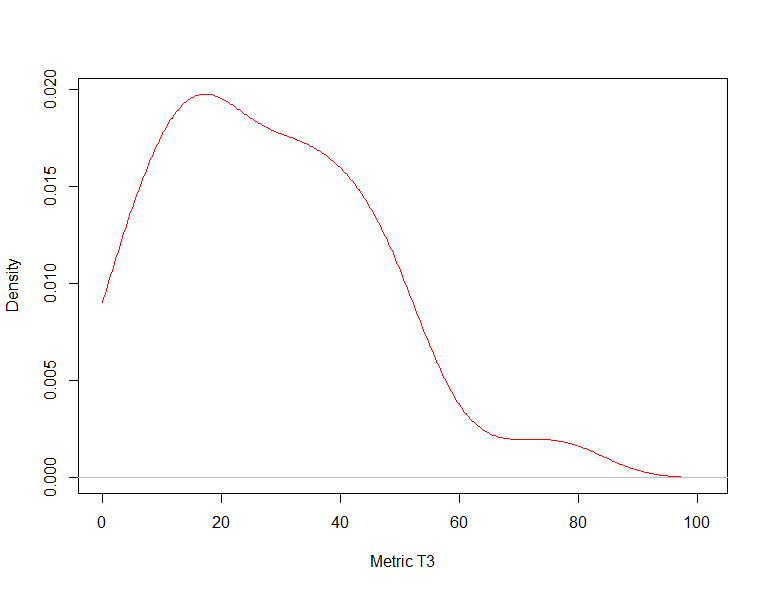}
         \end{subfigure}
              \begin{subfigure}[b]{0.48\textwidth}
         \centering
         \includegraphics[width=\textwidth]{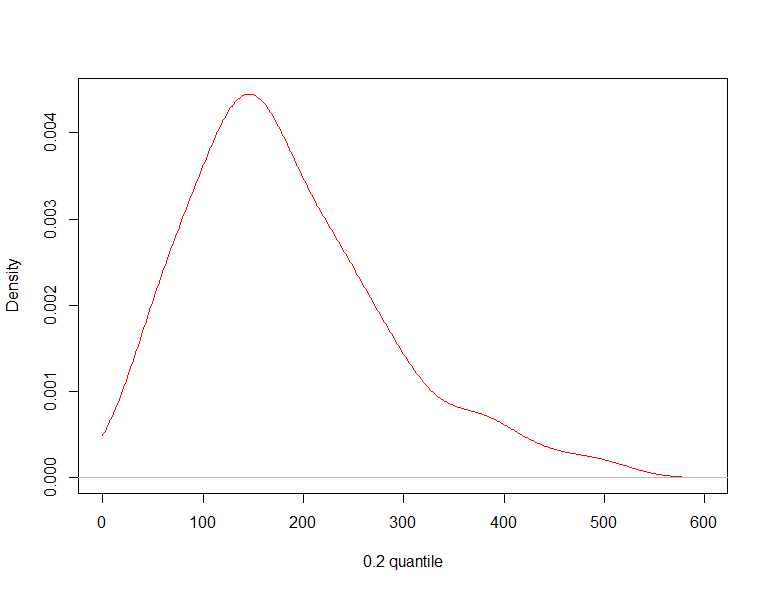}
         \end{subfigure}
              \begin{subfigure}[b]{0.48\textwidth}
         \centering
         \includegraphics[width=\textwidth]{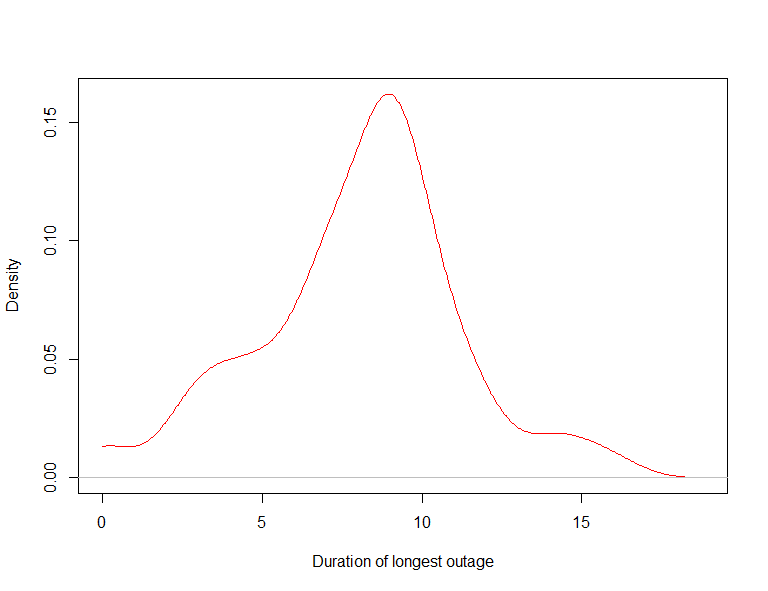}
     \end{subfigure}
    \caption{Kernel density estimators for $\tau$ metrics}
    \label{f2}
\end{figure}

As the conventional histogram can exhibit rather abrupt changes between discrete bins, it would be more advantageous to employ the kernel density estimator to reveal patterns in metric distributions. The results for these non-parametric estimations of the distributions of $\tau$ metrics are depicted in Figure~\ref{f2}. All the estimated distributions demonstrate unimodal behaviour, and most have a one-sided heavier tail. This indicates that the discussed metrics are suitable for characterising reliability based on these distributions.

\begin{figure}[!htb]
    \centering
    \includegraphics[width=0.95\linewidth]{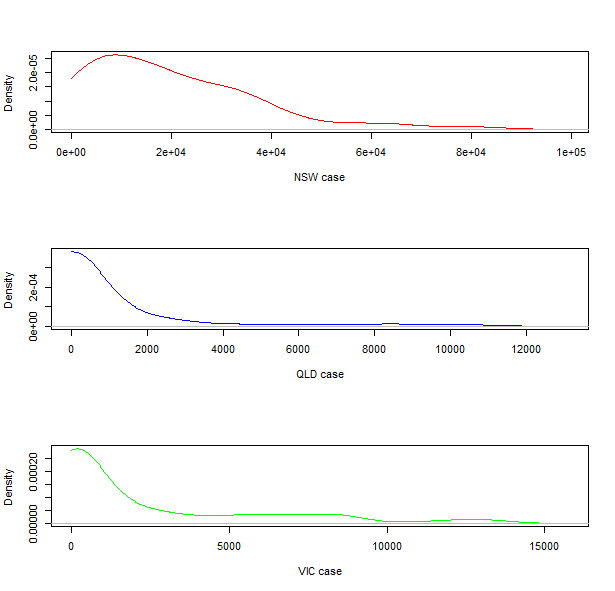}
    \caption{Distributions of metric T1 for 3 states}
    \label{f3}
\end{figure}

Figure~\ref{f3} illustrates the differences in scale and shape among the distributions of the metric T1 for the states of New South Wales, Queensland, and Victoria. The kernel density estimators for Queensland and Victoria display considerable similarity, with Victoria exhibiting a slightly heavier tail. In contrast, the kernel density estimator for New South Wales has a significantly larger main part.

\subsection{Sensitivity Analysis of Metrics to Length of Time Frame}

In  the previous numerical studies, we used the values from USE forecasts for the period spanning from 1/7/2025 to 30/6/2028. This entire time period served as an observation window for each of the 68  USE samples. Thus, it resulted in a total of 52,608 of 30 min time points for each sample.

One of the important practical questions is whether such long simulation or observation periods are required to obtain reliable estimators of the considered metrics. The issues pertaining to the sensitivity of statistical indices to the length of the observation period, the number of observations, or other factors are crucial in study design, selecting appropriate control parameters and ensuring the reliability of practical conclusions. For instance,  see the analysis of the sensitivity of statistical indices in \cite{Olen}.  This section explores the properties of the estimators for the reliability metrics with respect to the width of the observation window.

\begin{figure}[bh!]
    \centering
    \begin{subfigure}[b]{0.48\textwidth}
        \centering
        \includegraphics[width=\textwidth]{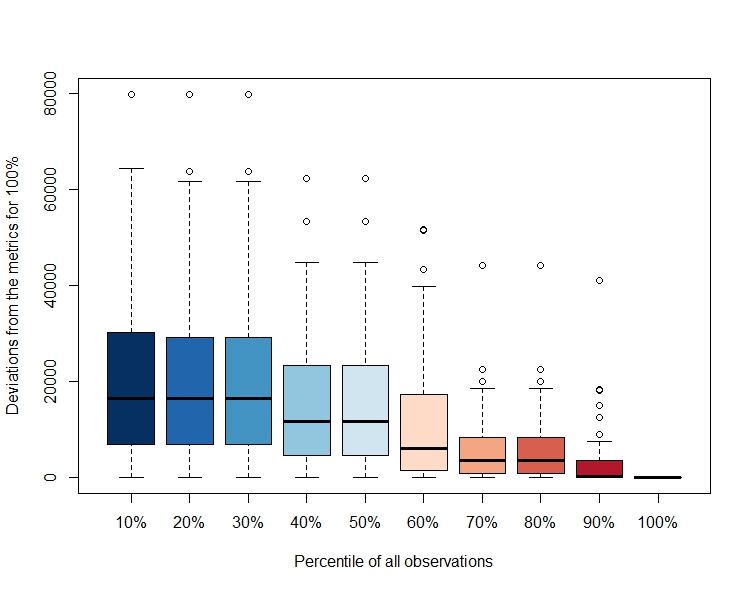}
    \end{subfigure}
    \begin{subfigure}[b]{0.48\textwidth}
        \centering
        \includegraphics[width=\textwidth]{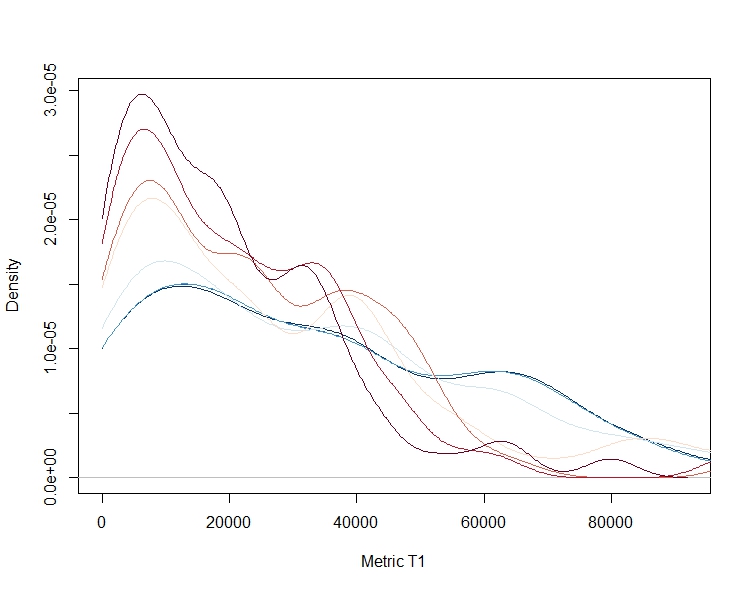}
    \end{subfigure}
    \begin{subfigure}[b]{0.48\textwidth}
        \centering
        \includegraphics[width=\textwidth]{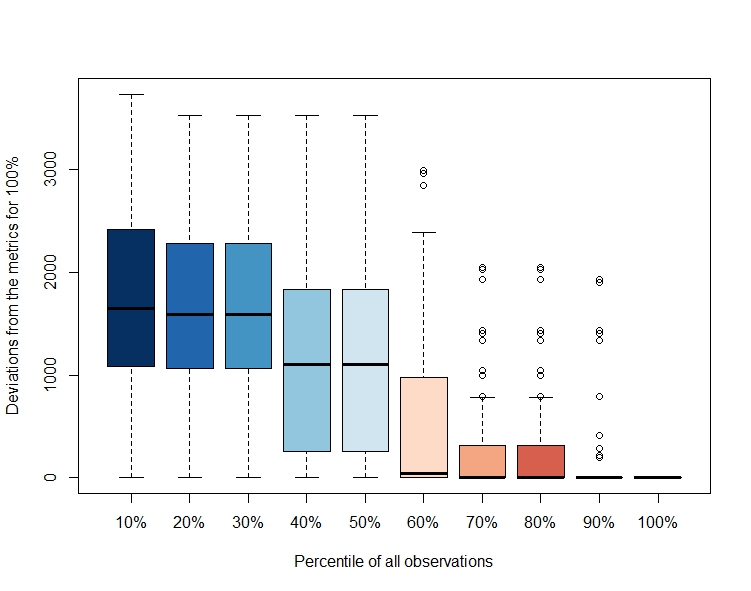}
    \end{subfigure}
    \begin{subfigure}[b]{0.48\textwidth}
        \centering
        \includegraphics[width=\textwidth]{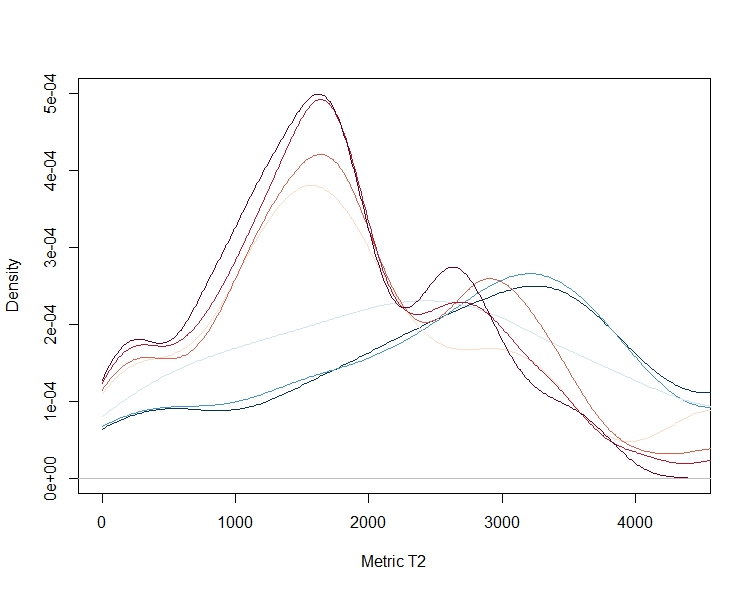}
    \end{subfigure}
    \begin{subfigure}[b]{0.48\textwidth}
    \centering
    \includegraphics[width=\textwidth]{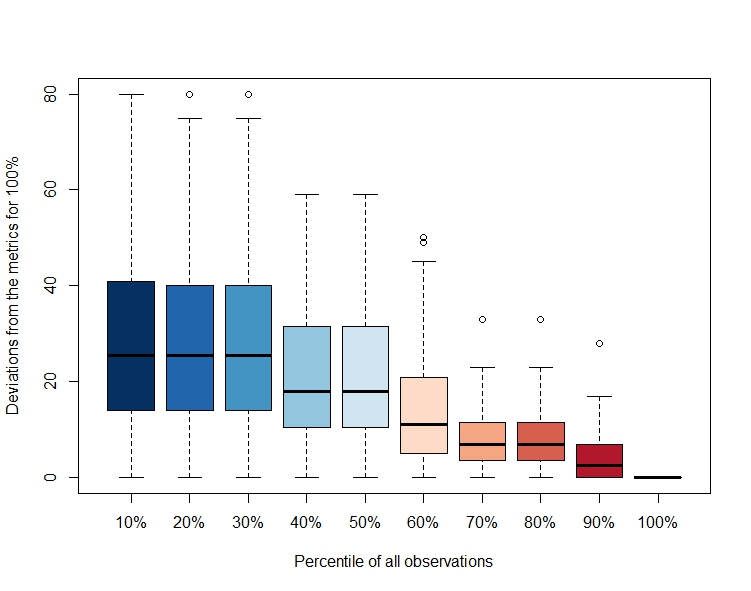}
\end{subfigure}
\begin{subfigure}[b]{0.48\textwidth}
    \centering
    \includegraphics[width=\textwidth]{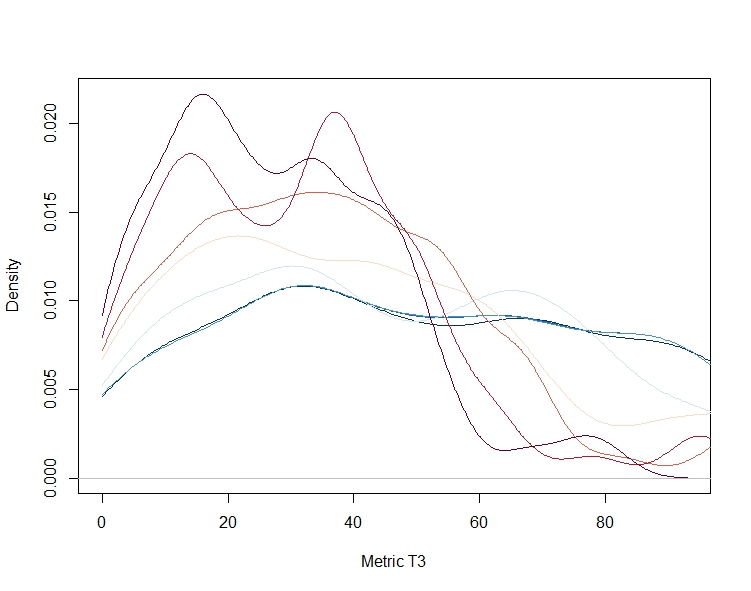}
\end{subfigure}
    \caption{Distributions of $\tau$ metrics depending on the width of estimation time frames}
\label{dev10}
\end{figure}

\begin{figure}[htb]
    \centering
    \begin{subfigure}[b]{0.48\textwidth}
    \centering
    \includegraphics[width=\textwidth]{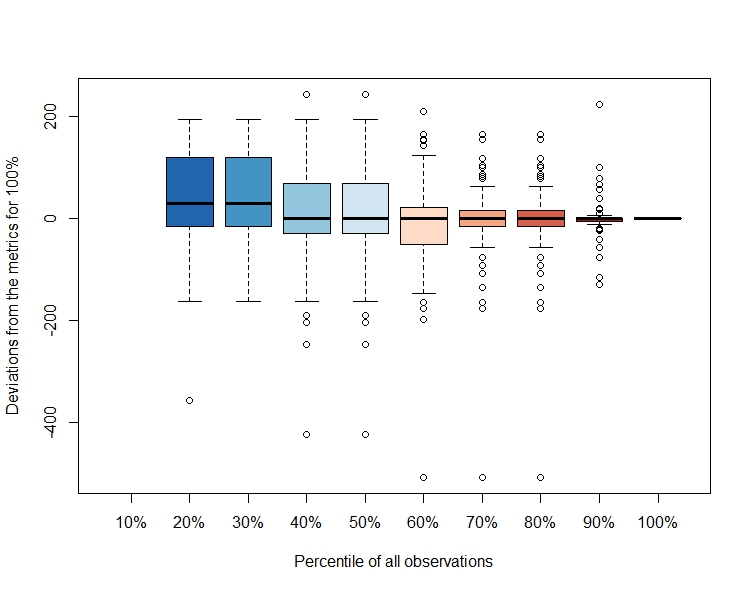}
\end{subfigure}
\begin{subfigure}[b]{0.48\textwidth}
    \centering
    \includegraphics[width=\textwidth]{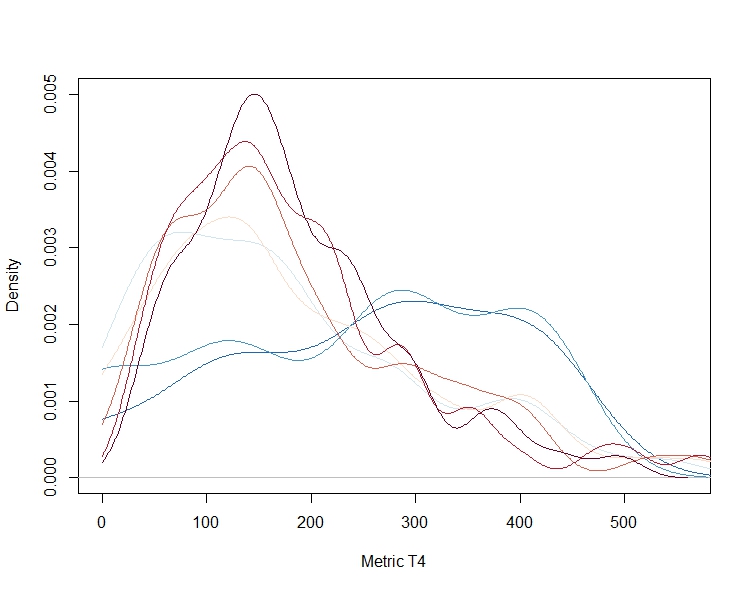}
\end{subfigure}
    \begin{subfigure}[b]{0.48\textwidth}
    \centering
    \includegraphics[width=\textwidth]{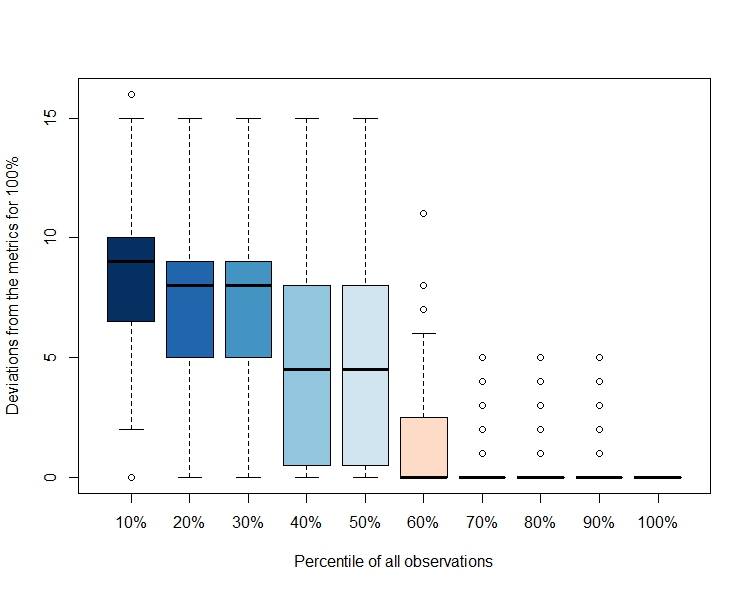}
\end{subfigure}
\begin{subfigure}[b]{0.48\textwidth}
    \centering
    \includegraphics[width=\textwidth]{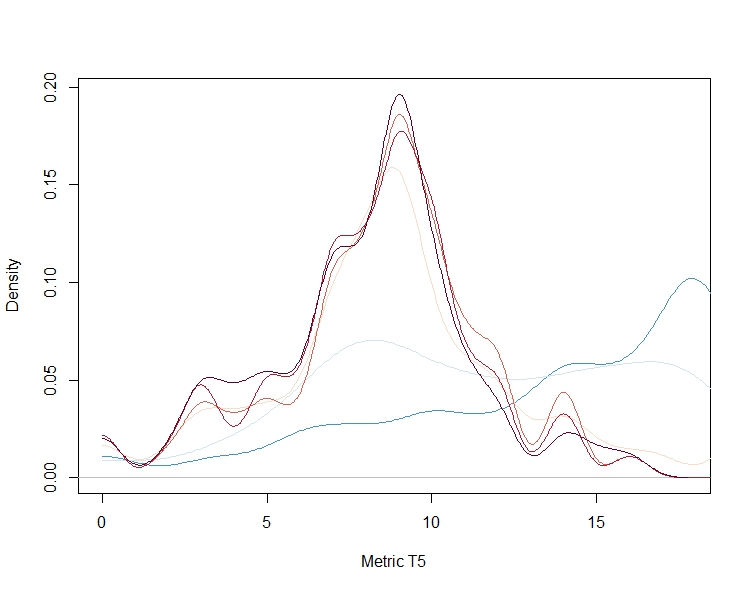}
\end{subfigure}
    \caption{Distributions of $\tau$ metrics depending on the width of estimation time frames}
    \label{dev11}
\end{figure}

We specifically examined observation windows that range from 10\% to 90\% of the total period. For each of these window lengths, we randomly selected a set of windows of the corresponding size and compute the values of each of the metrics under consideration. This process was repeated for each of 68 surveys. Subsequently, for the obtained values, we generated boxplots illustrating the deviations of the obtained sampled values of a metric from the corresponding metric values computed for the entire period. The obtained results for 5 metrics are depicted in the left hand side columns of Figures~\ref{dev10} and \ref{dev11}.


The right-hand side plots in Figures~\ref{dev10} and \ref{dev11} display the corresponding kernel-estimated densities derived from the computed values for each subwindow, ranging from 10\% to 100\%.Note that the boxplots for the 100\% subwindow serve as reference values, and, as expected, they display a value of zero. This result occurs because the length of the 100\% subwindow coincides with the total length of the observations.
Furthermore, it is worth mentioning that the density plots for the 100\% subwindow precisely mirror those in Figure~\ref{f2}. This congruence illustrates the consistency in the observed density distribution for this specific 100\% subwindows.

The data presented in Figures~\ref{dev10} and \ref{dev11} clearly indicate that shorter observation windows do not yield estimators of the same accuracy as the full duration of the observations. However, for all examined metrics, the accuracy of the estimators shows a significant improvement when the length of the observation window reaches 70\% of the total duration of simulation period. Additionally, a careful examination of the density plots suggests that such a window length offers a reasonable estimation of the distribution of the metrics.

Figures~\ref{dev10} and \ref{dev11} indicate that the convergence of the distributions for the third and fourth metrics to their respective limiting values occurs at a slower rate compared to that of other metrics. Furthermore, it appears that observation windows with a length comprising 50\% or less of the total window length do not yield reliable results. This inconsistency may lead to significantly different practical outcomes and distributions that deviate significantly from the expected limiting distributions.

Based on these numerical findings, it can be concluded that an observation window comprising 70\% of the total duration, which is equivalent to 36825 of 30min observations, can reliably provide appropriate estimates.

This  approach to investigate appropriate lengths of observation windows can be readily applied to any new metrics or their combinations, should the need arise.

\subsection{On the Dependency Structure of Selected\\ Metrics}
As several of the proposed metrics are non-linear functions of the observed data, their distributions become complex. Moreover, deriving explicit closed-form expressions for their joint distributions in the majority of cases would be very difficult or impossible. In this section, we use empirical estimates to illustrate this complexity.

\begin{figure}
    \centering
    \includegraphics[width=1.05\linewidth, trim =0.8cm 0 0 0, clip]{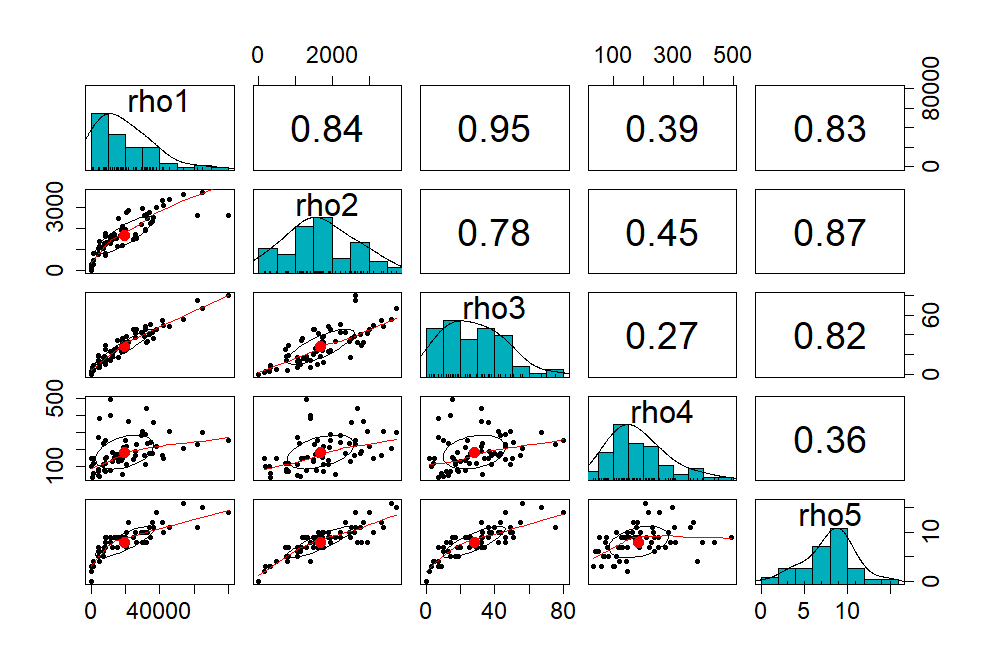}
    \caption{Scatter plots and correlations between the selected metrics. The diagonal gives a histogram with the fitted normal distribution for each metric}
    \label{dep1}
\end{figure}

Figure~\ref{dep1} depicts three types of information about the considered metrics: their bivariate scatter plots under the diagonal,  values of pairwise Pearson correlation coefficients above the diagonal, and their univariate histograms on the diagonal. First, upon visual inspection of the structure of dependencies using pairwise scatter plots between different metrics, it becomes evident that some metrics are closely related: see Figure~\ref{dep1}. For instance, metrics 1 and 3, as well as 2 and 5, exhibit strong correlations. However, metrics 4 and 2, as well as 4 and 3, do not demonstrate such relationships. Generally, metric 4 appears to be quite distinct from the other metrics under consideration. The empirical estimates of covariance functions above the diagonal validate these observations. The estimated correlation for the first-mentioned pairs is above 0.8, while for the second group of pairs, it falls below 0.45.

\begin{figure}[htb!]
    \centering
    \includegraphics[width=0.65\linewidth]{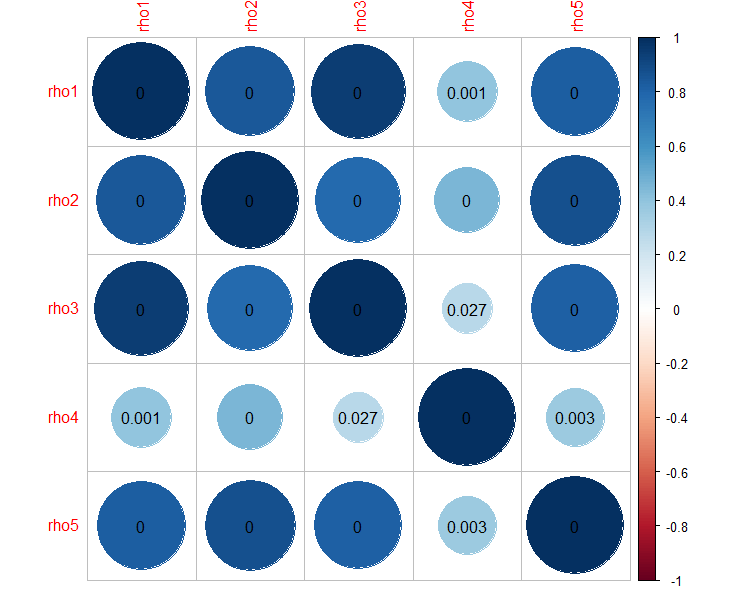}
    \caption{p-values of obtained correlations}
    \label{dep2}
\end{figure}

Figure~\ref{dep2} displays the p-values of the computed correlations. All of them are below the 0.05 threshold, indicating that the correlations of all metrics are significantly different from 0. As expected they are positively correlated.
When faced with the necessity of employing a set of metrics, information regarding the correlations can prove to be an important guideline in the selection of less correlated metrics. This specific selection will facilitate a more comprehensive characterisation of various properties pertaining to reliability characteristics.

\begin{figure}[htb!]
        \centering
        \includegraphics[width=0.9\linewidth]{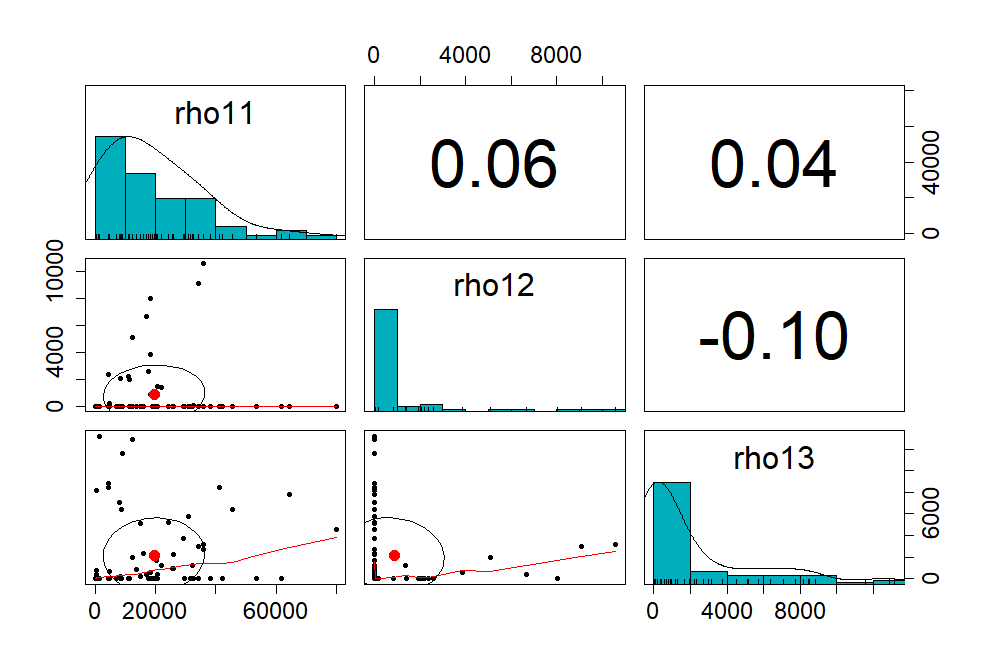}
    \caption{    Scatter plots and correlations of T1 values for different states. The diagonal displays histograms of T1 values with fitted normal distributions for each state.}
    \label{dep3}
\end{figure}

Finally, we compute the dependencies between the values of metric T1 for the three states depicted in Figure~\ref{f3}. The results displayed in Figure~\ref{dep3} demonstrate that, for this particular data set, there are no significant correlations among the values of this metric across states.

\section{Conclusion}\label{sec4}
The article presents a comprehensive proposal for new reliability standards that aim to ensure the resilience and reliability of the power sector in the face of the changing power system landscape. The proposed standards have been developed through a rigorous review process, taking into account the unique challenges and opportunities associated with integrating renewable energy sources and addressing resource adequacy. The paper proposes a comprehensive set of metrics and their combinations that effectively address reliability risks, balancing economic benefits and risk management, while also considering tail risks and providing a direct interpretation of outage severity.

By adopting these new reliability standards, policymakers, regulators, and industry stakeholders can enhance system planning, improve resource adequacy, and increase the resilience of the power sector. We hope that this article will serve as a valuable resource for decision-makers and contribute to the ongoing dialogue on enhancing the reliability of the National Electricity Market.

\subsection{Limitations}
It is important to acknowledge that the considered approach may have certain limitations.

These risk measures offer a straightforward assessment of the likelihood of service disruptions and can potentially help in determining  a consumer cost function through the use of surveys. However, the high dimensionality of some metrics could potentially impede optimisation in the integrated system plan.

Also, the proposed new reliability standards are based on the current understanding of the power system and the challenges it faces. However, as the power sector continues to evolve, new challenges and opportunities may arise that require further refinement and adaptation of the standards. 

Nonetheless, we believe that the proposed general framework can serve as a useful guidance and tool for developing different reliability standards in the future.

\subsection{Future Research}
To address these limitations, further research and analysis may be needed. Future research should focus on monitoring the implementation of the proposed standards, comparing and evaluating their effectiveness, and identifying areas for improvement. Additionally, ongoing research is necessary to stay abreast of technological advancements, market dynamics, and policy developments that may impact the reliability of the power system.

%\section*{Acknowledgements}
\paragraph{Acknowledgements}
We are grateful to the Australian Energy Market Commission for bringing the problem to MISG 2023 and to their representatives who attended the workshop to clarify various aspects of the problem. The industry representative, Craig Oakeshotte, along with moderators John Boland, David Hill, Ya Li, and Kihun Nam prepared the problem and materials and coordinated the activities of the MiISG group.

The organization and hospitality at Monash University was greatly appreciated.

We acknowledge the individuals listed below, in alphabetical order, who also contributed to this MISG group's discussions.

\bigskip
{\small
\noindent Joel Gilmore, Iberdrola Australia\\
Luke Gundry, School of Chemistry, Monash University\\
Ben Jones,  Australian Energy Market Operator\\
Taylor Kearney, Monash University\\
Daniel Uteda,  School of Mathematics and Statistics, the University of Melbourne\\
Jiahao Wu, School of Mathematics, Monash University}
	
% Reference
\bibliographystyle{abbrvnat}
\bibliography{misg.bib}

\section*{Authors Addresses}
\begin{enumerate} \item John Boland\\
UniSA STEM\\
The University of South Australia, Australia\\
email: John.Boland@unisa.edu.au\\
orcid: 0000-0003-1132-7589
\item Matthias Fresacher\\
School of Computer, Data and Mathematical Sciences\\
Western Sydney University, Australia\\
email: M.Fresacher@westernsydney.edu.au\\
orcid: 0000-0003-0677-3701
\item David Hill\\
Department of Electrical and Computer Systems Engineering\\ Monash University, Australia\\
email: DavidJ.Hill@monash.eduu\\
orcid: 0000-0003-4036-0839
\item  Shijia Jin\\
School of Mathematics\\
Monash University, Australia\\
email: shijia.jin@monash.edu
\item  Ya Li\\
School of Mathematics\\
Monash University, Australia\\
email: ya.li@monash.edu
\item  Graham Mills\\
Australian Energy Market Commission\\
email: Graham.Mills@aemc.gov.au
\item   Kihun Nam\\
School of Mathematics\\
Monash University, Australia\\
email: kihun.nam@monash.edu\\
orcid: 0000-0002-6570-3861
\item   Craig Oakeshotte\\
Australian Energy Market Commission\\
email: Craig.Oakeshott@aemc.gov.au
\item   Andriy Olenko\\
Department of Mathematical and Physical Sciences\\
La Trobe University, Australia\\
email: a.olenko@latrobe.edu.au\\
orcid: 0000-0002-0917-7000
   \end{enumerate}
\end{document}